# Entropy is a Mathematical Formula

Jozsef Garai
E-mail: jozsef.garai@fiu.edu

**Abstract**

The microscopic explanation of entropy has been challenged from both experimental and theoretical point of view. The expression of entropy is derived from the first law of thermodynamics indicating that entropy or the second law of thermodynamics is not an independent law.

## Introduction

The macroscopic determination of entropy first was expressed by Clausius in 1865. He postulated that the entropy [S] change between two equilibrium states could be determined by the transferred reversible heat [Q] and the absolute temperature [T] as:

$$dS = \frac{dQ}{T} \tag{1}$$

Employing statistical mechanics in 1877 Boltzmann suggested a microscopic explanation for entropy. He stated that every spontaneous change in nature tends to occur in the direction of higher disorder, and entropy is the measure of that disorder. From the size of disorder entropy can be calculated as:

$$S = k_B \ln W \tag{2}$$

where W is the number of microstates permissible at the same energy level, $k_B$ is the Boltzmann constant. The modification of the Boltzmann's expression has been proposed recently by Tsallis for non equilibrium systems[1-3]. The conventional entropy expression can be applied when a system is in thermal equilibrium; however, for nonextensive systems the following expression is proposed.

$$S_{Tsallis} = k_B \frac{W^{1-q} - 1}{1-q} \tag{3}$$

When parameter q goes to 1 then equation 3 gives equation 2, which is the Boltzmann solution[4].

**Experimental problems**

The microscopic explanation of entropy has never been fully accepted[5-7] since there is incomplete proof for equation 2 and there are counter examples where the increase of disorder cannot be justified. Among these counter examples the spontaneous crystallization of a super-cooled melt[8] and the crystallization of a supersaturated solution[9] are the most prominent challenges for the Boltzmann's model. If a super-cooled melt is allowed to crystallize under adiabatic conditions then the entropy of the system increases. In a supersaturated solution there is a possibility of the deposition of crystalline solute. The deposition of crystalline solute is a spontaneous process with an increase of entropy. A crystallization of a liquid or a deposition of a crystalline solute does not imply any increase in disorder. I would like to add one additional examples to the above mentioned. The phase transformation of solid helium to liquid helium II does not require energy[10]. The lack of latent heat results the same entropy for both of these phases [Fig. 1], indicating the same level of disorder in solid and in liquid. These counter examples suggesting that the microscopic explanation of entropy is dubious.

**Theoretical problems**

The number of microstates for a given macrostate can be deduced from a combinatorial argument[11] and can be calculated as

$$W = \frac{N!}{n_0! n_1! n_2! n_3! \ldots} \tag{4}$$

where N is the total number of particles and n is the number of particles at the same energy level. The total number of particles and the total energy [E] can be written as

$$N = \sum_{j=0}^{\infty} n_j \quad \text{and} \quad E = \sum_{j=0}^{\infty} n_j E_j \tag{5}$$

where $E_j$ is the energy of the jth level.

$$E_j = j\varepsilon \tag{6}$$

where $\varepsilon$ is the energy difference between to consecutive states. As an example the number of microstates in the possible macrostates is given in Figure 2. The system consist 4 particles with the total energy of $E = 6\varepsilon$. Macrostate with the highest number of microstates is the most probable; however, macrostates with lower numbers for the microstates are also possible but their probability is lower. According to this model the number of microstates can vary while the total energy remains the same. Allowing to change the number of microstates without affecting the total energy of the system is in discrepancy with equation 1 and 2.

## The formula of entropy

Assuming a quasi-static infinitesimal process and employing the first law of thermodynamics, the differential of the internal energy dU is the sum of the heat term dq and the mechanical work term dw.

$$dU = dq + dw \tag{7}$$

At constant volume the work is zero.

$$dU_V = dq_V = nc_V dT \tag{8}$$

where n is the number of moles, and $c_V$ is the molar heat capacity at constant volume. Assuming a monoatomic gas, where only translational energies are present the constant molar volume heat capacity is

$$c_V = \frac{3}{2}R \tag{9}$$

where R is the universal gas constant. Substituting R with $\frac{pV}{nT}$ from the equation of state [EoS] for ideal gasses:

$$dU_V = \frac{3}{2}pV\frac{dT}{T} \tag{10}$$

where V is volume and p is the pressure. Integrating the equation gives

$$\Delta U_V = \frac{3}{2}pV\int_{T_i}^{T_f}\frac{dT}{T} = \frac{3}{2}pV\ln\left(\frac{T_f}{T_i}\right) \tag{11}$$

where subscript i represent the initial conditions while f represents the final conditions. Employing EoS and substituting pV with nRT and 1.5R with $c_V$

$$\Delta U_V = nc_V T \ln\left(\frac{T_f}{T_i}\right) \tag{12}$$

This equation contains the expression of entropy for constant volume.

$$\Delta S_V = nc_V \ln\left(\frac{T_f}{T_i}\right) \tag{13}$$

The internal energy then

$$\Delta U_V = \Delta q_V = T\Delta S_V \tag{14}$$

At constant temperature the heat term is zero and equation 7 is written.

$$dU_T = dw = -pdV \tag{15}$$

Substituting p with $\frac{nRT}{V}$ gives:

$$dU_T = dw = -nRT\frac{dV}{V} \tag{16}$$

Integrating the equation between the initial and final conditions:

$$\Delta U_T = -nRT\int_{V_i}^{V_f}\frac{dV}{V} = -nRT\ln\left(\frac{V_f}{V_i}\right) \tag{17}$$

Equation 17 contains the expression of entropy for constant temperature.

$$\Delta S_T = nR\ln\left(\frac{V_f}{V_i}\right) \tag{18}$$

Substituting entropy into equation 17 gives

$$\Delta U_T = \Delta w = -T\Delta S_T \tag{19}$$

In a general case, when both the temperature and the volume change, the internal energies should be summed.

$$\Delta U = \Delta q_V + \Delta w = \Delta U_V + \Delta U_T = T\Delta S_V - T\Delta S_T = T(\Delta S_V - \Delta S_T) \tag{20}$$

Since the heat transferred to the system supplies the energy for both the heat and the work the sign of the work changes.

$$\Delta Q = \Delta q_V - \Delta w = T\Delta S_V - (-T\Delta S_T) = T(\Delta S_V + \Delta S_T) \tag{21}$$

Summing the two parts of the entropy change

$$\Delta S = \Delta S_V + \Delta S_T \tag{22}$$

gives the heat

$$\Delta Q = T\Delta S \tag{23}$$

This equation is the same as equation 1 proposed by Clausius.

## Conclusion

The microscopic explanation of entropy, that is the entropy measures the disorder of a system, should be discredited since contradict with experiments and theory. The expression of entropy can be derived from the first law of thermodynamics suggesting that the second law of thermodynamics is not an independent law.

The formula for entropy most likely was introduced to provide a convenient way to calculate the changes in the internal energy of a system and to convert the thermal and mechanical energies into each other.

## References

(1) Tsallis, C. *J. Stat. Phys.* **1988** 52, 479.
(2) Curado, E.M.F.; Tsallis, C. *J. Phys. A* **1991** 24, L69.
(3) For complete references, http://tsallis.cat.cbpf.br/biblio.html
(4) Cho, A. *Science* **2002** 297, 1268.
(5) Wright, P. G. *Contemp. Phys.* **1970** 11, 581.
(6) Dingle, H. *Bull. Inst. Phys.* **1959** 10, 218.
(7) Khinchin, A. I. *Mathematical Foundations of Statistical Mechanics*; English translation: New York, 1949.
(8) Bridgman, P. W. *The Nature of Thermodynamics*; Cambridge, Mass., 1941.
(9) McGlashan, M. L. *J. Chem. Educ.* **1966** 43, 226.
(10) Swenson, C.A. *Phys. Rev,* **1950** 79, 626.
(11) Schoepf, D.C. *Am. J. Phys.* **2002** 70, 128.

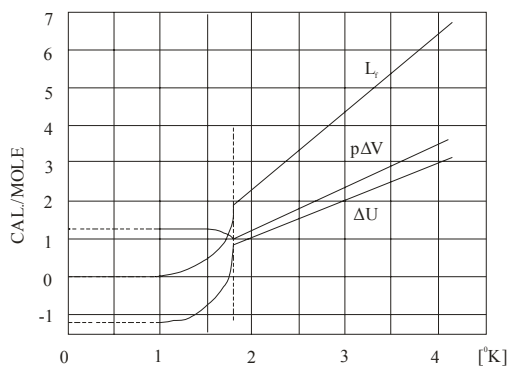

**Figure 1.** The molar latent heat of fusion $L_f$, $p\Delta V$, and the change in internal energy, $\Delta U = L_f - p\Delta V$ for helium from Ref. 10. The lack of latent heat results the same entropy for these phases suggesting the same disorder for solid and liquid phase.

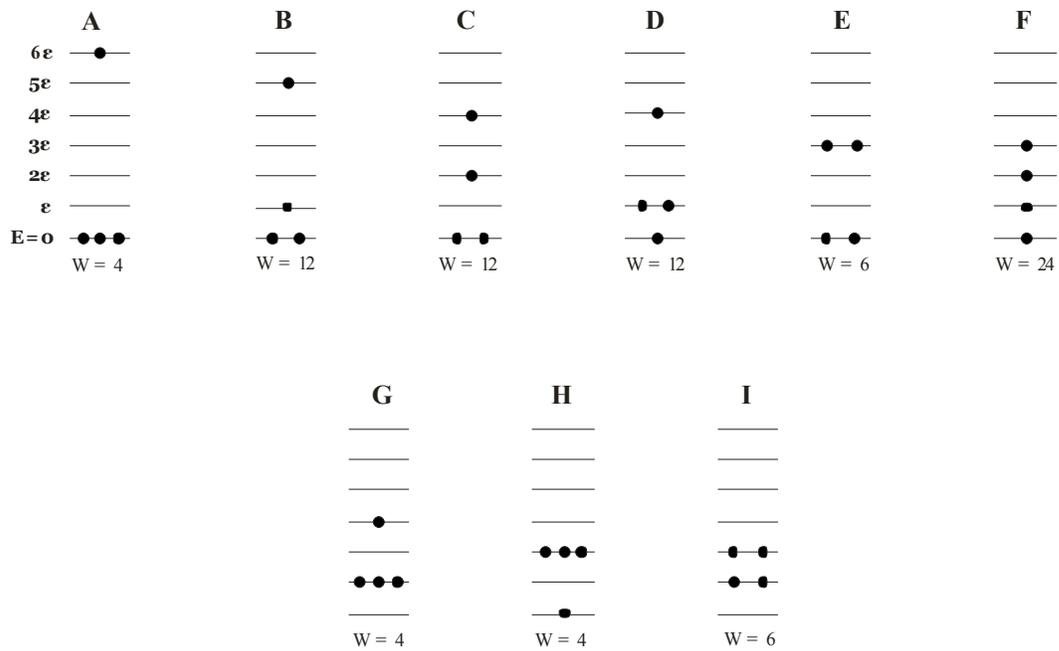

**Figure 2.** The possible macrostates for a system of N = 4 particles with total energy E = 6ε and their number of microstates from Ref 11.